\documentstyle[rotate,epsf,psfig]{l-aa}
\thesaurus{ 06(02.16.1 ; 02.16.2; 02.18.7 ; 02.23.1 ; 08.16.6 ) }

\begin{document}

\title{Natural Polarization Modes in Pulsar Magnetospheres}

\author{ A. ~von Hoensbroech \inst{1,2},
H. ~Lesch \inst{2}
and T. ~Kunzl \inst{2,3}}

\offprints {A.~von~Hoensbroech (avh@mpifr-bonn.mpg.de)}

\institute{Max-Planck-Institut f\"ur Radioastronomie,
 Auf dem H\"ugel 69, D-53121 Bonn, Germany.
\and 
Institut f\"ur Astronomie und Astrophysik der
 Universit\"at M\"unchen, Scheinerstr. 1, D-81679
 M\"unchen, Germany.
\and
Max-Planck-Institut f\"ur extraterrestrische Physik, 
Giessenbachstr., D-85740 Garching, Germany.
}

\date{ Received 9. March 1998 / Accepted 28. April 1998}

\maketitle
 
\markboth{~von Hoensbroech et al.: Natural Polarization Modes in 
Radio Pulsar Magnetospheres}{}

\begin{abstract}

We present a comprehensive investigation
of the radio polarization properties within the theory
of natural electromagnetic wave modes in pulsar
magnetospheres. Taking into account the curvature of the field
lines, aberration effects and magnetic sweep-back we use
the relativistic dielectric tensor in the low-density cold plasma 
approximation
and derive the following polarization characteristics,
which are in full agreement with the observational findings.
Specifically, we demonstrate that 
{\bf 1.} The degree of linear polarization decreases with increasing
frequency. 
{\bf 2.} The degree of circular polarization increases with increasing
frequency.
{\bf 3.} At high frequencies ($\geq$ a few GHz) the degree of linear 
polarization is correlated to the spin down luminosity $\dot E$.
{\bf 4.} At high frequencies long-period pulsars exhibit weaker
linear polarization than their short-period counterparts.
{\bf 5.} The difference between the refractive indices of the two 
natural wave modes decreases with increasing frequency which 
possibly results into a depolarization via superposition.

\end{abstract}

  
\section{Introduction}
\label{intro}

The polarization properties of pulsar radio emission are highly 
diverse. Nearly all possible
polarization states have been detected, from pulsar to pulsar, from 
frequency to frequency, from pulse to pulse and even within
one pulse profile. Nevertheless, some general statements can be made
from an observational point-of-view, which must be explained by any 
plausible theoretical model:

\begin{itemize}
\item Pulsar radio emission is usually highly linearly polarized at low
frequencies (in the following ``low frequencies'' defines frequencies
up to about one GHz, ``high frequencies'' corresponds to everything
well above one GHz). Towards higher frequencies the radiation tends to 
depolarize (e.g. \cite{M71}, \cite{X96}).
\item The observed polarization position angle (hereafter PPA)
is a very stable feature which usually does not depend strongly on
frequency. Apart from the occurrence of sudden $\sim90^\circ$ jumps, 
the so-called orthogonal polarization modes (hereafter OPM), the PPA
usually follows a characteristic S-shaped curve. This curve can 
well be modelled by the rotating vector model (hereafter RVM) 
as proposed by Radhakrishnan \& Cooke (1969). It is therefore 
thought to reflect
the geometry of the radiating field lines.
\item The sum of many (typically at least a few hundred) single 
pulses usually converges towards a characteristic average
profile. In contrast, the individual pulses show a high variability. 
One well known single pulse feature is the {\it quasi-periodic 
microsecond-structure} (e.g. \cite{CH77}, \cite{L98}) observed over
a broad frequency interval with no obvious frequency dependence
of the quasi-periodicity. It is usually highly linearly and 
circularly polarized with a roughly constant PPA and often OPMs 
at the edges. This suggests that, during one micro-pulse, only one
of the orthogonal polarization modes is observed. 
Another feature are the {\it subpulses}, 
sometimes drifting in pulse phase from one pulse to the next one. 
The PPA usually shows a swing during one subpulse, with the swing 
being stable relative to the subpulse. So when the subpulse drifts, 
the swing drifts as well, causing a depolarization in the integrated 
profile (e.g. \cite{MTH75}).
\item In average profiles and in individual pulses OPMs can be 
observed (e.g. Stinebring et al. 1984, Gangadhara 1997). 
The magnitude of these jumps is 
usually $90^\circ $ but can also be less sometimes.
Sudden orthogonal jumps in the 
PPA can be found in nearly all pulse phases and even at the same
pulse phase in successive single pulses. The OPMs often coincide
with changes in the handedness of the circular polarization. Their
existence therefore suggests that the radiation preferentially takes
two orthogonal elliptical states.
\item Some pulsars show a high degree of circular polarization, 
preferentially near the centroid of the profile. Sometimes it also sometimes 
shows a sense reversal near the centroid (\cite{RR90}). Recent 
observations show that a few pulsars have a strong increase in their
degree of circular polarization towards high frequencies together
with a decrease in linear polarization, thus suggesting that a 
process might be active which transforms linear into circular
polarization (\cite{HKK98}). 
\item At low frequencies there seems to be no significant relation
between the average degree of polarization and any other pulsar
parameter. At higher frequencies, however, correlations have been
found between the degree of polarization and the spin down luminosity
$\dot E$ respectively the surface acceleration potential (\cite{M81}, 
\cite{X95} and \cite{HKK98}). Pulsars with a high $\dot E$
(respectively a short period and a large period derivative) show
a higher average polarization at high frequencies than those with 
a lower $\dot E$. This suggests that for these pulsars a possible 
depolarization process becomes important at higher frequencies
than for the other pulsars.
Closely connected to this is an anti-correlation between
the degree of polarization at high frequencies and the period. Only
rapidly rotating pulsars seem to have highly polarized radio emission
(\cite{HKK98}).
\item Several authors estimated the emission altitudes where
the observed radiation is produced and studied if there is a frequency
dependence of this altitude (Cordes 1978, Rankin 1990, Thorsett 1991,
Blaskiewicz et al. 1991, Gil 1991, Phillips 1992, Gil \& Kijak 1997,
von Hoensbroech \& Xilouris 1997, Kramer et al. 1997).
Various
independent methods were used and all authors came to the conclusion
that the radio emission originates from a region close to the pulsar
surface at a few percent of the light cylinder radius 
(at least for ``normal'' -- non millisecond --  pulsars). The frequency
dependence of the emission height is thought to be small, if it exists
at all.

\end{itemize}

All these observational facts have to be explained
within a general model for the polarization of pulsar radio emission.
Basically, there are two aspects which must be considered.
One is the polarization characteristic of the emission process
itself, and the other one is the influence of the magnetosphere
on the propagating radiation. In this paper we concentrate on the
possible role of propagation for the polarization because this 
can be considered independently of the possible
radiation mechanisms.

The magnetosphere of a pulsar is filled with a plasma which is 
streaming relativistically outwards along the open 
field lines. 
A number of articles has already been published on propagational
effects in this plasma.

Cocke \& Pacholczyk (1976) considered the
effect of Faraday pulsation for quasi-transverse propagation. Faraday
pulsation is the general case of Faraday rotation. It occurs when a 
polarized beam decays into two elliptical orthogonal modes which 
propagate with different phase velocities. The PPA of the resulting 
beam gets rotated and the polarization changes between linear and
circular. 

\noindent
In a series of papers Melrose \& Stoneham (1977), 
Melrose (1979) and Allen \& Melrose (1982) derived an approximation for
the dispersion tensor and calculated the properties of the natural 
wave modes in the plasma (see also Lyutikov 1998). They assumed that 
the polarization of the 
propagating electromagnetic wave follows the shape of the natural
wave modes up to a certain radius of limiting polarization (hereafter
$R_{\rm LP}$). 
The radiation then escapes with the polarization properties
of the natural wave modes at $R_{\rm LP}$.
Barnard (1986) places $R_{\rm LP}$ at the place of cyclotron resonance. 
For normal pulsars this resonance occurs at a few 10\% of the light
cylinder radius, but for short period pulsars ($P\leq 0.06 s$) this resonance 
is outside $R_{\rm LC}$.

\noindent
Cheng \& Ruderman (1979)  considered adiabatic walking to be responsible for
the variable polarization in subpulses. The polarization properties
were derived for different emission mechanisms and plasma
compositions.

\noindent
Harding \& Tademaru (1979, 1980, 1981) presented numerical calculations
on the propagation of linearly polarized pulses through a shearing
plasma. They show that this can account for micro-structure, 
circular polarization and rotation of the PPA.

\noindent
Barnard \& Arons (1986) and Arons \& Barnard (1986) have calculated 
the dispersion relations
for the X- and the O-mode in an ultra-relativistic, one-dimensional 
plasma for distribution functions in $e^-$ and $e^+$ (the O-mode
has its electric field vector in the plane of curvature of the
magnetic field, the X-mode perpendicular to it). The X-mode
is purely linearly polarized and propagates easily through
the magnetosphere whereas the O-mode can have some circular 
contributions and follows the bending of the field lines. This
causes a separation of the OPMs (\cite{McK97}).

\noindent
Calculations concerning the propagation characteristics of the 
various natural modes have also been presented by 
Beskin et al. (1993). Using the limit of infinite magnetic field 
strength they conclude that only the electromagnetic X-mode can 
propagate freely through the magnetosphere. The angular dependence
of the properties of the modes and the influence of the 
cyclotron resonance are derived. This resonance
becomes only significant, if the secondary particle production rate is 
high ($\ge 10^4$ per primary particle) thus leading to a much higher
plasma density. The different mode properties also account 
for the circular polarization often observed in core components.

In this paper we extended the ideas proposed by 
Melrose \& Stoneham(1977), Melrose (1979) and Allen \& Melrose (1982).
We derive the properties of the natural wave modes throughout 
the magnetosphere with special respect to the angle between the 
propagating wave and the magnetic field. 
We make predictions for the qualitative dependence of the polarimetric 
properties with frequency and various pulsar parameters. These predictions 
are then compared with the observations.

\section{Method}

\subsection{Global picture}\label{assumptions}
For our calculations we use the following global picture for the
pulsar radio emission (e.g. Manchester \& Taylor 1977): 
The neutron star (hereafter NS) has a strong dipolar magnetic field 
which produces a
corotating magnetosphere filled with charged particles which are
drawn from the NS surface. 
It is convenient to define the light cylinder radius $R_{\rm LC}$ 
of the magnetosphere at which the corotation velocity is equal to the 
speed of light
\begin{equation}
R_{\rm LC}=\frac{cP}{2\pi}\ ,
\end{equation}
where $c$ is the speed of light and $P$ is the rotation period of the NS.
Those field lines which do not close within $R_{\rm LC}$ define the
polar cap. The angular size of the polar cap is therefore given
by
\begin{equation}\label{lof}
\phi_{\rm lof}=\arcsin\left(\sqrt{\frac{R_{\rm NS}}{R_{\rm LC}}}\right)
\end{equation}
with $R_{\rm NS}$ being the neutron star radius ($R_{\rm NS}=10\,\rm km$,
lof stands for {\bf l}ast {\bf o}pen {\bf f}ield line).
The particles get accelerated above the polar cap and flow
relativistically outwards along the open field lines. 
The motion is one dimensional (all particles are in their lowest
Landau-levels). Only one type of particles is considered
(here electrons). Some pair producing processes might be active, thus
adding another type of particles but we assume that there will always 
be one type dominating. The problems of strong pair production follows
from considerations by Kunzl et al. which will be presented in
a forthcoming paper. 

According to the model of Goldreich \& Julian (1969) we
use the following density:
\begin{equation}\label{nGJ}
n_{\rm GJ}=\frac{4\pi\epsilon_0}{e}\cdot\frac{B_0}{P}
\end{equation}
with $e$ being the elementary charge and 
$\epsilon_0=8.85\cdot10^{-12}{\rm AsV^{-1}m^{-1}}$ 
the electric field constant. The surface magnetic field
is defined as
\begin{equation}\label{B0}
B_0=\sqrt{\frac{3c^3\mu_0 I}{32\pi^3 R_{\rm NS}^6\sin^2\alpha}\cdot P\dot P}\ .
\end{equation}
Here $I=10^{38} {\rm kg}\cdot {\rm m}^2$ is the moment of inertia of the NS 
and $\alpha$ is the angle between the axis of
rotation and the magnetic field axis. $\dot P$ is the temporal 
derivative of the period, $\mu_0=4\pi\cdot 10^{-7} {\rm VsA^{-1}m^{-1}}$ 
the magnetic field constant.
As the field is dipolar, the magnetic field strength and the local plasma 
density decrease with the cube of the distance from the NS.

At a few percent of $R_{\rm LC}$ the plasma
produces radio emission. Since we want to make our calculations
independent of the actual emission mechanism, we only assume
that the particles initially have a 
certain Lorentz factor $\gamma_{\rm beam}$, just before they radiate. The
angle of emission is approximately $1/\gamma_{\rm beam}$ with respect to 
the field line along which they stream.
After having radiated the particles flow further outwards and form the {\it
background} plasma (bg) through which the radio waves have to
propagate. Naturally the background plasma has a lower Lorentz factor 
than the radiating beam, i.e. $\gamma_{\rm
bg}\ll\gamma_{\rm beam}$. Since the radio emission mechanism must 
be coherent and therefore 
very efficient, we assume that the energy spread of the background 
plasma is negligible above the emission region, thus the plasma is
cold. In this case the 
cold plasma approximation applies and 
the energy distribution function is 
\begin{equation}\label{cold}
f(\gamma)\simeq\delta(\gamma -\gamma_{\rm bg}). 
\end{equation}

Further, we assume that we observe only one natural wave mode
at a time. The polarization properties of the wave correspond to the
local form of the natural wave modes up to $R_{\rm LP}$.
Above this radius, the interaction between the propagating wave and the
plasma becomes so weak that the escaping wave preserves the state of 
polarization of that radius to infinity (see e.g. Melrose 1979, 
Barnard 1986 and Beskin et al. 1993).
The maximum value for $R_{\rm LP}$ is $R_{\rm LC}$ but for
most pulsars $R_{\rm LP}$ is probably reached at a closer distance.
The minimum value for $R_{\rm LP}$ is the emission height itself.

The assumption that only one polarization mode is observed at a time 
might not always be true, but the existence of
highly polarized OPMs indicates that this is at least sometimes true. 
If both modes are 
observed simultaneously the radiation gets depolarized.
The results of the following discussion can therefore
be regarded as upper limits for the observed degrees of polarization.

\subsection{Geometrical Aspects}\label{geo}

It is very important to analyze carefully the angle of 
the propagating wave relative to the magnetic field. 
This can be understood by an
inspection of the extreme cases for parallel and transverse
propagation. For an electromagnetic wave propagating parallel to
a magnetic field, the natural wave modes are circular (R- and L-mode),
whereas 
for transverse propagation they are linear (X- and O-mode)
(see e.g. \cite{KT86} or any standard plasma physics book). 
For oblique propagation intermediate states appear (see
Fig. \ref{trans}). It is our aim 
to calculate these states as a function of the various parameters 
involved.

To derive the angle between the propagating wave and the
magnetic field lines the following contributions have to be
considered:
The angle which originates from the dipolar structure of the
field, the angle associated with the {\it aberration} and the degree of 
{\it magnetic sweep back}. 
Aberration is the apparent forward bending of the field lines through the 
rotation of the pulsar (\cite{P92}). The magnetic sweep back is 
caused by the magnetic dipole radiation. The magnetic field lines are 
bent in the direction opposite to the pulsar rotation by the
electromagnetic torque of the neutron star (\cite{S83}).

We do not consider the relativistic
beaming angle ($\theta_{\rm beaming}=1/\gamma_{\rm beam}$) 
outside the emitting region itself because this angle 
is symmetrical with respect to the direction of emission and will 
be averaged out statistically.

\subsubsection{Dipolar contribution}

\begin{figure}
\vspace{.5cm}
\epsfysize5cm
\rotate[r]{\epsffile[150 200 450 500]{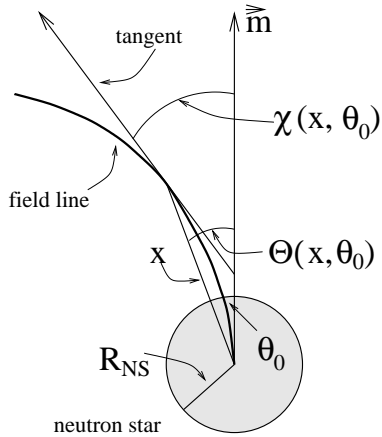}}
\caption[]{Geometry to derive the tangential inclination
$\chi(x,\theta_0)$ of any dipolar field line with the footpoint-angle
$\theta_0$ at the radial distance $x$.}
\label{dipol1}
\end{figure}

In a polar coordinate system, the quotient $\sin^2\theta/r$ is
constant for a dipolar field line. Having the NS in the centre, and
defining each field line by its {\it footpoint} (point of
intersection) at the
NS-surface ($R_{\rm NS}$,$\theta_0$), a parametric equation describing the
field line is given by:
\begin{equation}\label{pardp}
\frac{\sin^2\theta}{r}={\rm
constant}\equiv\frac{\sin^2\theta_0}{R_{\rm NS}}\ .
\end{equation}
Some straightforward calculations lead to an expression for the
tangential angle $\chi(x,\theta_0)$ of the $\theta_0$-field line:
\begin{equation}
\chi(x,\theta_0)=\Theta(x,\theta_0)+\arctan\left(\frac{1}{2}\tan 
\Theta(x,\theta_0)\right)\; .
\end{equation}
Here $x=r/R_{\rm NS}$ and 
$\Theta(x,\theta_0)=\arcsin(\sqrt{x}\cdot\sin\theta_0)$ is
the radial angle of the $\theta_0$-field line at a distance $x$
(see also Lyne \& Graham-Smith 1990).
The geometry is displayed in Fig. \ref{dipol1}.
This can be used to calculate the polar angle $\theta_{\rm S}$ for a
given distance $x$
at which a ray propagates, that was emitted tangentially from the 
$\theta_0$-field line at the emission height $x_{\rm em}$:
\begin{eqnarray}
\theta_{\rm S}(x,\theta_0)&=&\chi(x_{\rm em},\theta_0)-\arcsin\left[
\frac{x_{\rm em}}{x}\cdot\sin\left(\pi-\right.\right.\nonumber\\
&-&\left.\left.\chi(x_{\rm em},\theta_0)+\Theta(x_{\rm
em},\theta_0)\right)\rule[-.25cm]{0pt}{.65cm}
\right]\ .
\end{eqnarray}
Using this point ($x$,$\theta_{\rm S}$), the footpoint-angle 
$\theta_{\rm S0}$ of the 
corresponding field line can be calculated using Eq. (\ref{pardp}).

\begin{figure}
\vspace{.5cm}
\epsfysize5cm
\rotate[r]{\epsffile[80 150 480 450]{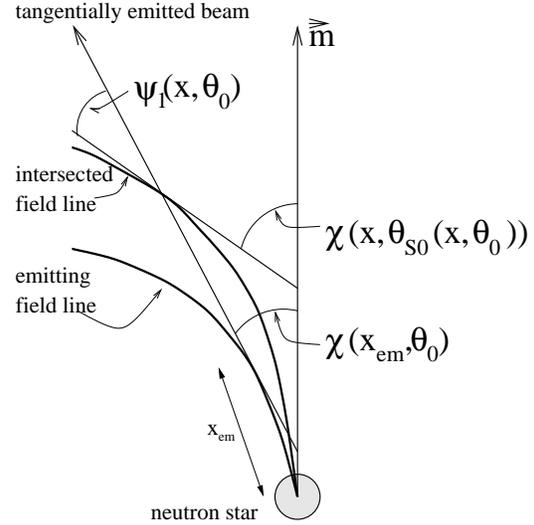}}
\caption[]{Geometry to derive the intersecting angle
$\psi_1(x,\theta_0)$ between a beam which was emitted tangentially
from the $\theta_0$-field line at the emission height $x_{\rm em}$ and
a field line in a radial distance $x$.}
\label{dipol2}
\end{figure}

The intersecting angle between a ray emitted tangentially 
from the $\theta_0$ field line in the emission height $x_{\rm em}$ with
the field line at the distance $x$ can then be found as the difference 
of the tangential angles
\begin{equation}
\psi_1(x,\theta_0)=\chi(x,\theta_{\rm S0}(x,\theta_0))-
\chi(x_{\rm em},\theta_0)\ ,
\end{equation}
as is shown in Fig. \ref{dipol2}.

\subsubsection{Other contributions}

The rotation of the magnetosphere causes additional effects which have
to be considered for the intersecting angle.
The radiation is bent forward into the direction of rotation by {\it
aberration}. Following Phillips (1992), the magnitude of this effect is 
\begin{equation}
\theta_{\rm ab}(x)=\arctan\left( 2\pi\frac{x R_{\rm
NS}\sin\alpha}{c P}\right)\ .
\end{equation}

At larger distances from the NS, corotation of the plasma cannot be 
maintained, and {\it magnetic sweep-back} becomes important. The
calculation of this effect is rather complex, but an approximation
was given by Shitov (1983):
\begin{equation}
\theta_{\rm msb}(x) \simeq 1.2\cdot\left(\frac{2\pi x R_{\rm NS}}{c
P}\right)^3 \cdot \sin^2\alpha\ .
\end{equation}
The direction of this effect is backwards relative to the rotation.
To calculate the net-effect of these two contributions, one has to
take their difference at the distance $x$ and subtract their value at
the emission height $x_{\rm em}$. It is therefore given by
\begin{equation}
\psi_2(x)=\theta_{\rm ab}(x)-\theta_{\rm ab}(x_{\rm em})
-\theta_{\rm msb}(x)+\theta_{\rm msb}(x_{\rm em})\ .
\end{equation}
Angular contributions by effects of general relativity
are negligible at distances larger than a few km from the
NS surface (e.g. Gonthier \& Harding 1994). Since the contribution
will always be much smaller than the accuracy of our 
calculations, which is restricted to the beaming angle of 
$1/\gamma_{\rm beam}$.

\subsubsection{Resultant intersection angle}\label{resang}

\begin{figure}
\vspace{.5cm}
\epsfysize5cm
\rotate[r]{\epsffile[80 150 560 450]{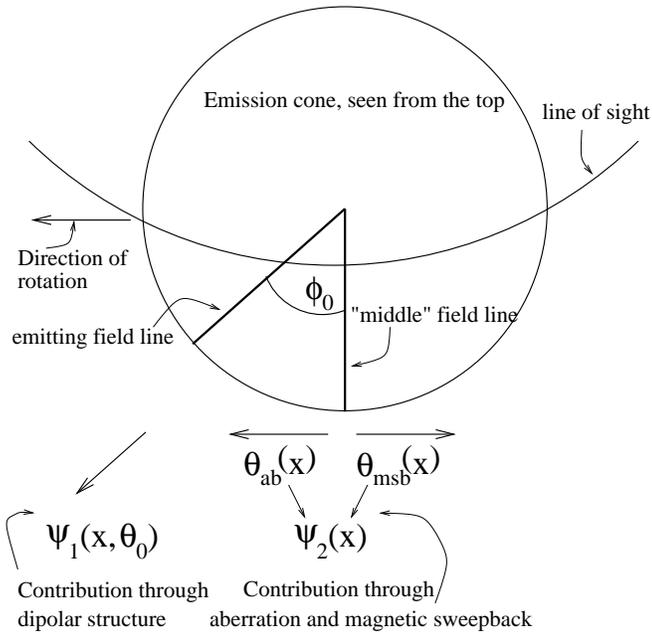}}
\caption[]{Geometry to derive the resultant intersection angle under
consideration of the dipolar contribution and those of aberration and
magnetic sweep-back. $\psi_1$ points towards the direction of the emitting
field line, $\theta_{\rm ab}$ into the direction of
rotation. $\theta_{\rm msb}$ points backwards relative to the
direction of rotation.}
\label{dipol3}
\end{figure}

For the resultant intersection angle (see Fig. \ref{dipol3}), the
curvature direction of the
emitting field line has to be taken into account. This can be defined
as the angle between the plane of curvature of the emitting field line
and the field line in the middle of the profile $\phi_0$. 
The tube of emitting field lines then rotates into the direction
of the field line with $\phi_0=90^\circ$. 

The resultant angle of a beam, radiated from the
$\theta_0$-field line into the direction $\phi_0$ at the emission
height $x_{\rm em}$, which intersects with a field line in a distance
$x$ from the NS can then be given as:
\begin{eqnarray}
\psi_{\rm S}(x,\theta_0)&=&\left[(\psi_1(x,\theta_0) \cos\phi_0)^2+
\right.\nonumber\\
&+&\left.(\psi_2(x)-\psi_1(x,\theta_0)\sin\phi_0)^2\right]^\frac{1}{2}\
.
\end{eqnarray}

This equation is only valid if one regards the intersecting angle at a
sufficient distance from the emitting altitude, where $\psi_{\rm S}$
reaches a finite value. At the emission region itself, $\psi_{\rm S}$
would be zero. As the radiation is beamed into a forward-cone of
$1/\gamma_{\rm beam}$, and the intersected field lines are all nearly
parallel in the emission region, we use this beaming-angle 
$\psi_{\rm S}=1/\gamma_{\rm beam}$ for $x\simeq x_{\rm em}$.

\begin{figure}
\vspace{.5cm}
\epsfysize5cm
\rotate[r]{\epsffile[120 50 520 450]{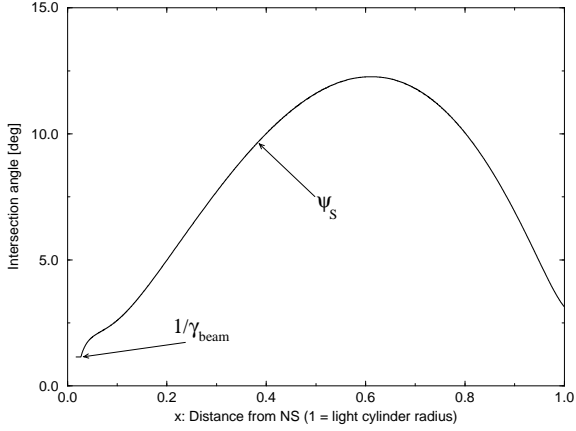}}
\caption[]{Resultant intersection angle for a canonical pulsar (see text
for assumed parameters). The angle is displayed from the emission height
to the light cylinder radius.}
\label{winkel}
\end{figure}

In Fig. \ref{winkel}, the intersection angle between a tangentially
emitted beam and a field line at a given distance is displayed 
from the emission height to $R_{\rm LC}$ for a canonical pulsar. 

\subsubsection{The canonical pulsar}\label{canonical}

As we require a number of intrinsic pulsar parameters for our calculations,
we use the mean values from the known pulsar sample.
These parameters are $P=0.6\, {\rm s}$ and $\dot
P=10^{-14.6}$. For the magnetic inclination we assume
$\alpha=40^\circ$. The regarded field line is at $\phi_0=40^\circ$ 
and $\theta_0=\theta_{\rm lof}/1.5$ ($\theta_{\rm lof}$ is the footpoint
angle of the last open field line, see Eq. \ref{lof}). As a 
typical emission height we choose $2\%$ of $R_{\rm LC}$, which
corresponds to slightly more than 50 pulsar radii for the canonical pulsar. 
For the Lorentz factor of the background plasma we use be 
$\gamma_{\rm bg}=1.7$, following Kunzl (1997). In those cases, where
we derive properties at a given distance in the magnetosphere, we 
use a point at $20\%$ of the light cylinder 
radius for $R_{\rm LP}$ (see also Sect. \ref{qual}). 
As a typical value for $\gamma_{\rm beam}$
we assume a value of 50 (e.g. \cite{Rn96}, \cite{Kz97}). As stated earlier, 
the value of $\gamma_{\rm beam}$ is only important very close to the emission 
region itself.

\subsection{Approximation for the Dielectric Tensor}

In order to derive the properties of the natural wave modes it is
necessary to get an approximate idea of the dielectric characteristics
of the plasma. Our approach follows the work of Melrose (1979), 
which is based on two assumptions:
\begin{itemize}
\item The plasma is in the {\it low density limit}, which is defined
by the condition that the Alfv$\acute{\rm e}$n  velocity is much
larger than the velocity of light ($v_{\rm A} \gg c$). Combining
the Alfv$\acute{\rm e}$n velocity
\begin{equation}
v_{\rm A}^2=\frac{B^2}{\mu_0\rho}
\end{equation}
with the relativistic mass density $\rho=\gamma_{\rm bg}m_{\rm e}\cdot
n$  leads to the condition 
\begin{equation}
\frac{B^2}{c^2\cdot\mu_0\gamma m_{\rm e}n}\gg1\ .
\end{equation}
This condition is well fulfilled throughout the pulsar magnetosphere. 
For our calculations this implies that the dielectric tensor is close
to unity. 
\item The background-plasma is assumed to be cold. This follows from 
the consideration that the coherent radiation mechanism can only be 
effective for particle energies above a certain threshold (see Eq. \ref{cold}).
The final
particle energies will therefore peak at this critical value. 

Thus the {\it long wavelength
approximation} applies, which can be used when the wavelength 
of the propagating
wave is much larger than the relevant lengths in the plasma. 
This approximation should be fulfilled throughout in the pulsar 
magnetosphere above the emitting region.
\end{itemize}

Using the low density limit the dielectric tensor may be written as
\begin{equation}
\epsilon_{ij}=\delta_{ij}+\Delta\epsilon_{ij}
\end{equation}
with $\Delta\epsilon_{ij}\ll1$. The dispersion equation is then found
as 
\begin{equation}
An^4-Bn^2+C=0
\end{equation}
with $A,B$ and $C$ being functions of $\epsilon_{ij}$. In this
approximation $\Delta\epsilon_{ij}$ does not depend on $n$. Hence two
solutions can be given for $n^2$ (\cite{MS77}):
\begin{eqnarray}\label{brecheq}
n^2&=&1\pm\Delta n_\pm^2\\
\Delta n_\pm^2&=&\frac{a+b\pm\sqrt{(a-b)^2+4g^2}}{2}\ .
\end{eqnarray}
Using the cold plasma approximation for electrons the dielectric
tensor leads to
\begin{eqnarray}
a&=&-\frac{\omega_{\rm pe}^2(\cos\theta-\beta)^2}{\omega^2(1
-\beta\cos\theta)^2 -\Omega_{\rm e}^2}-\frac{\omega_{\rm
pe}^2\sin^2\theta}{\gamma_{\rm bg}^2\omega^2(1-\beta\cos\theta)^2}\\
b&=&-\frac{\omega_{\rm pe}^2(1-\beta\cos\theta)^2} 
{\omega^2(1-\beta\cos\theta)^2 -\Omega_{\rm e}^2}\\
g&=&\frac{\omega_{\rm pe}^2(\Omega_{\rm e}/\omega)(\cos\theta-\beta)}
{\omega^2(1-\beta\cos\theta)^2 -\Omega_{\rm e}^2}\ ,
\end{eqnarray}
$\beta=v/c$. The electron plasma frequency $\omega_{\rm pe}$ and the
electron gyro frequency $\Omega_{\rm e}$ are defined as:
\begin{eqnarray}
\omega_{\rm pe}&\equiv&\sqrt{\frac{n\cdot e^2}{\gamma m_{\rm e}\epsilon_0}}\\
\Omega_{\rm e}&\equiv&\frac{|e|\cdot B}{\gamma m_{\rm e}}\ .
\end{eqnarray}
$n$ is the plasma density ($n(x)=n_{\rm GJ}/x^3$, see Eq. \ref{nGJ}), $e$
the elementary charge, $m_{\rm e}$ the electron mass and $\epsilon_0$ the
electric field constant. $B$ is the magnetic field strength 
($B(x)=B_0/x^3$, see Eq. \ref{B0}).

With these parameters, the axial ratio of the polarization ellipses
of the two natural wave modes can be calculated:
\begin{equation}
T_\pm=\frac{2g}{a-b\mp\sqrt{(a-b)^2+4g^2}}\ .
\end{equation}
$T=0$ and $T=\pm\infty$ correspond to linear polarizations. For elliptical
polarizations $T$ becomes finite. The rotating sense of a positive $T$
is right hand with respect to the direction of propagation and left hand
for negative $T$.

The degree of circular polarization can then be calculated through
\begin{equation}
\Pi_{{\rm V}\pm}:=\sin(2\arctan T_\pm)\;  .
\end{equation}
As a natural wave mode is fully polarized per definition, the linear
polarization can be calculated as the quadratic complement:
\begin{equation}\label{qucompl}
\Pi_{{\rm L}\pm}:=\sqrt{1-\Pi_{{\rm V}\pm}^2}\; .
\end{equation}

\section{Results}

The polarization of the natural wave modes depends mainly on the angle
between the propagating electromagnetic wave and the magnetic 
field. If the wave propagates purely parallel to the field ($\vec k
\parallel \vec B$, $\vec k$ being the wave vector of the propagating
beam), the modes obviously have to be circularly polarized since no
linear direction perpendicular to $\vec k$ is preferred. When $\vec k$ is
perpendicular to $\vec B$, a direction is preferred and the wave is
fully linearly polarized. If the angle between $\vec k$ and $\vec B$
becomes larger than $90^\circ$ then the handedness of the circular
polarization gets reversed for each mode due to the opposite direction 
of the particle gyration.

\begin{figure}
\vspace{.5cm}
\epsfysize5cm
\rotate[r]{\epsffile[120 50 520 450]{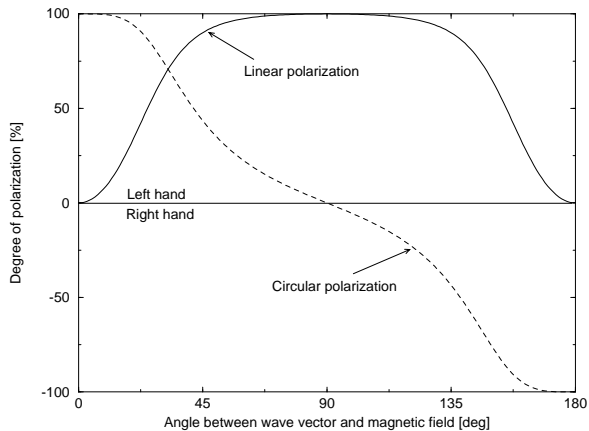}}
\caption[]{Degree of polarization of one natural wave mode
versus the intersecting angle between the propagating wave and the
magnetic field. In this figure we show the principal polarization
behaviour of 
one natural mode of a wave propagating at an oblique angle to the magnetic
field through a non relativistic ($\gamma_{\rm bg}=1$) cold  electron
plasma.  The function was calculated for a frequency of 1 GHz and 
a magnetic field strength of 0.2 T. For small angles between 
$\vec k$ and $\vec B$ the
polarization is mainly circular, as no linear direction is preferred
perpendicular to $\vec k$. At an angle close to $90^\circ$ the
polarization is mainly linear, for angles larger than $90^\circ$ it
becomes increasingly circular again but with an reversed handedness
(due to the opposite direction of gyration).
The transition between fully circular to fully linear
depends heavily on the frequency and the magnetic field strength. For 
plasma Lorentz factors $\gamma_{\rm bg}>1$, the angle gets Lorentz
transformed -- thus shifting the condition for transverse 
propagation to smaller angles.}
\label{trans}
\end{figure}

The transition (for an example see Fig. \ref{trans}) between these 
two states depends strongly on a number of parameters:
\begin{itemize}
\item The magnetic field strength. For a strong magnetic field, the
transition from circular to linear polarization 
occurs for very small angles between $\vec k$ and $\vec B$.
\item The frequency. For lower frequencies, a higher degree of linear
polarization is reached at smaller angles than for higher frequencies.
\item The plasma $\gamma$-factor. For a relativistic plasma, the
intersecting angle between the propagating beam and the magnetic field
is Lorentz-transformed, such that the angle increases. The
transverse propagation ($\sim 90^\circ$ in Fig. \ref{trans}) is reached
at a smaller angle.
\end{itemize}

\subsection{Qualitative polarization characteristics}\label{qual}

We apply these calculations to the propagation of an electromagnetic 
wave through
the pulsar magnetosphere using canonical pulsar parameters 
(Sect. \ref{canonical}) and taking into account all angular effects.

\begin{figure}
\vspace{.5cm}
\epsfysize5cm
\rotate[r]{\epsffile[120 50 520 450]{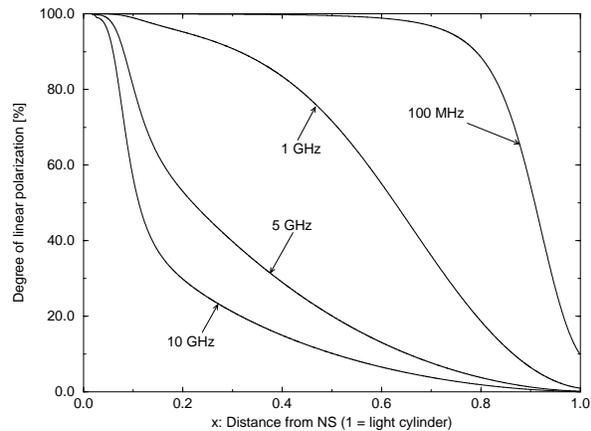}}
\caption[]{Degree of linear polarization of one natural wave mode
versus the distance from the NS (up to the light cylinder radius) 
for the canonical pulsar (Sect. \ref{canonical}).}
\label{lvsx}
\end{figure}

Figure \ref{lvsx} illustrates the change of the natural wave mode
polarization characteristics with distance from the pulsar for
different frequencies. For clarity, we displayed only the linear 
polarization
for one mode (the + mode, see Eq. \ref{brecheq}). 
As the circular polarization is given by
the quadratic complement (see Eq. \ref{qucompl}), a decrease of the
linear polarization leads to an increase of the circular
contribution. The dominating parameter is the rapidly decreasing 
magnetic field. After the radiation, the kinetic energy of the plasma
is relatively low. Following Kunzl (1997), we use a background Lorentz 
factor of $\gamma_{\rm bg}=1.7$. This value would result from a free
electron maser emission mechanism through comparing bunching- and loss
times. Other radiation mechanisms lead to different values of
$\gamma_{\rm bg}$ but the qualitative manner does not change. The
dependence on $\gamma_{\rm bg}$ is discussed separately in
Sect. \ref{lorentz}. 

The following qualitative behaviour can be observed.
Close to the emitting region, the mode is fully linearly polarized
for all frequencies due to the extremely high magnetic field in the
emitting region ($|\vec B|\simeq 10^3$ T). As the wave propagates 
outwards, the intersecting angle increases and the magnetic field
decreases. Initially this just causes the linear polarization of the high
frequency waves to decrease, while the lower frequency waves maintain a
high degree of linear polarization up to a large distance from the NS.
Thus, if only one mode is observed at a time, for a given distance we 
expect a {\it decrease} of linear polarization
with frequency and an {\it increase} of circular polarization. 

As stated earlier, it remains unclear up to which 
distance from the NS the polarization of the propagating radiation 
is influenced by the changes of the natural wave modes ($R_{\rm LP}$). 
This remains to be a problem to
be solved by mode coupling theory but appears to be rather
complicated as the plasma is most likely highly inhomogeneous in
space and time. Therefore for the present discussion we choose an arbitrary 
distance for $R_{\rm LP}$ and derive the qualitative behaviour without loss
of generality.

\begin{figure}
\vspace{.5cm}
\epsfysize5cm
\rotate[r]{\epsffile[120 50 520 450]{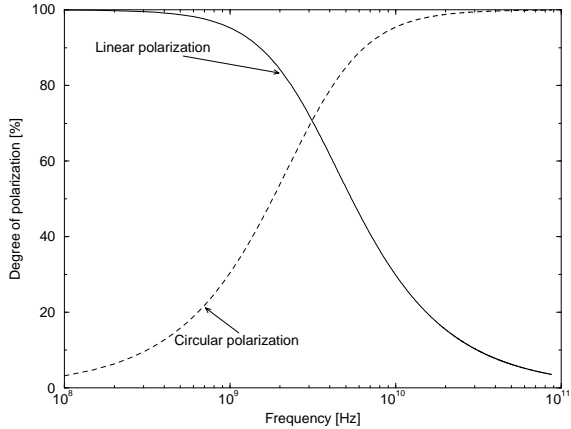}}
\caption[]{Degree of polarization of one natural wave mode
versus frequency. The height above the NS was chosen to be 20\% of the
light cylinder without loss of generality. The qualitative behaviour
is a decrease of linear and an increase of circular polarization with
frequency for the canonical pulsar (Sect. \ref{canonical}).}
\label{freq1}
\end{figure}

Figure \ref{freq1} shows the frequency development of one natural wave mode
at an arbitrary distance of 20\% of $R_{\rm LC}$ for a
typical pulsar. The degree of linear polarization should decrease and the
degree of circular should increase above about one GHz.

\subsection{Dependence on various pulsar parameters}

We now want to investigate, how the change of polarization
characteristics with frequency depends on the variation of other
pulsar parameters. 

\subsubsection{Spin-down luminosity}

\begin{figure}
\vspace{.5cm}
\epsfysize5cm
\rotate[r]{\epsffile[120 50 520 450]{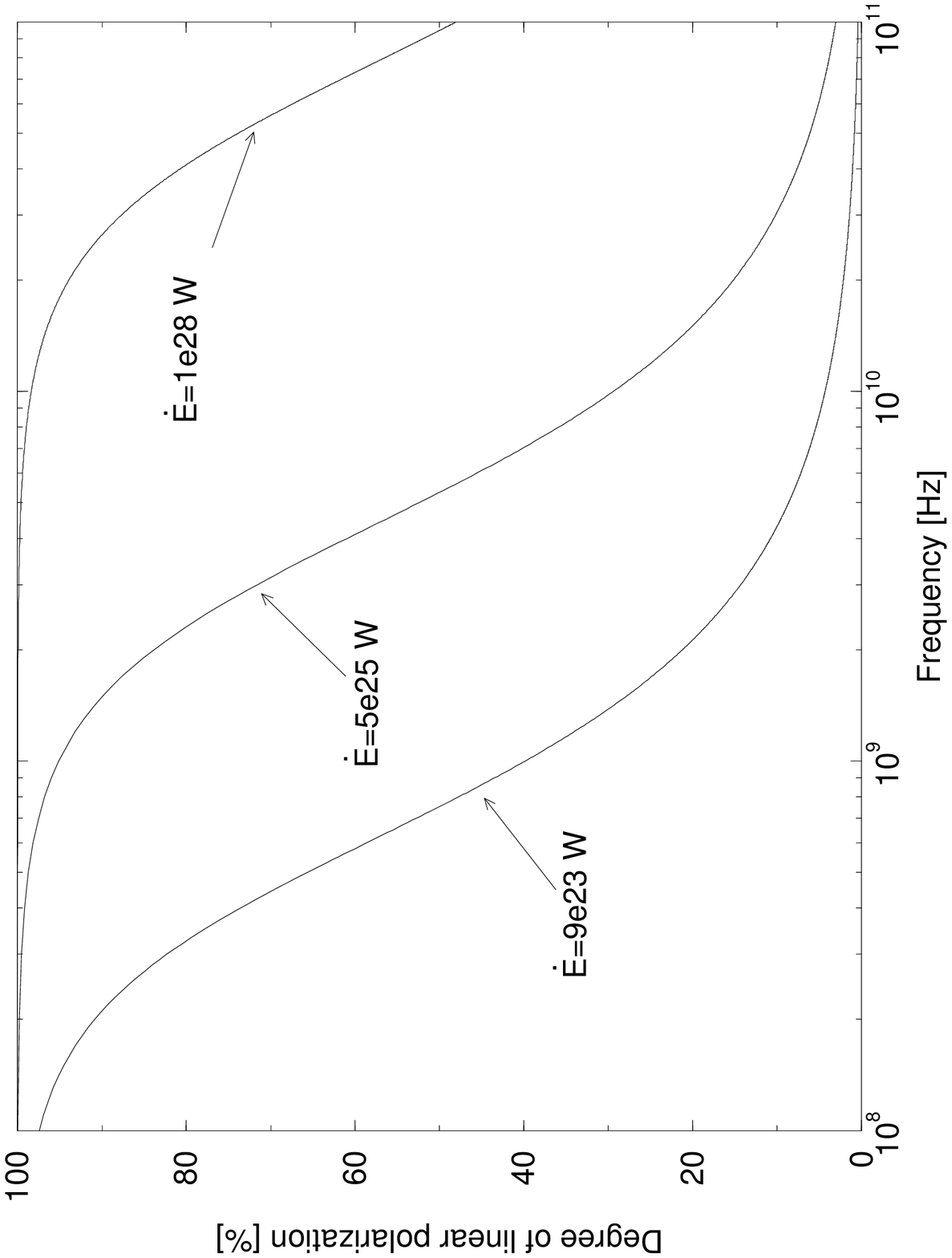}}
\caption[]{Degree of linear polarization of one natural wave mode
versus frequency for different spin down luminosities $\dot E$. We
used the same canonical pulsar parameters as before except that we
varied the values for
$\dot P$, keeping the period fixed at $P=0.6$ s. The resulting values
for $\dot E$ are written next to the curves. For pulsars with a 
higher $\dot E$
the decrease in linear polarization occurs at a higher frequency.
}
\label{edot}

\vspace{.5cm}
\epsfysize5cm
\rotate[r]{\epsffile[120 50 520 450]{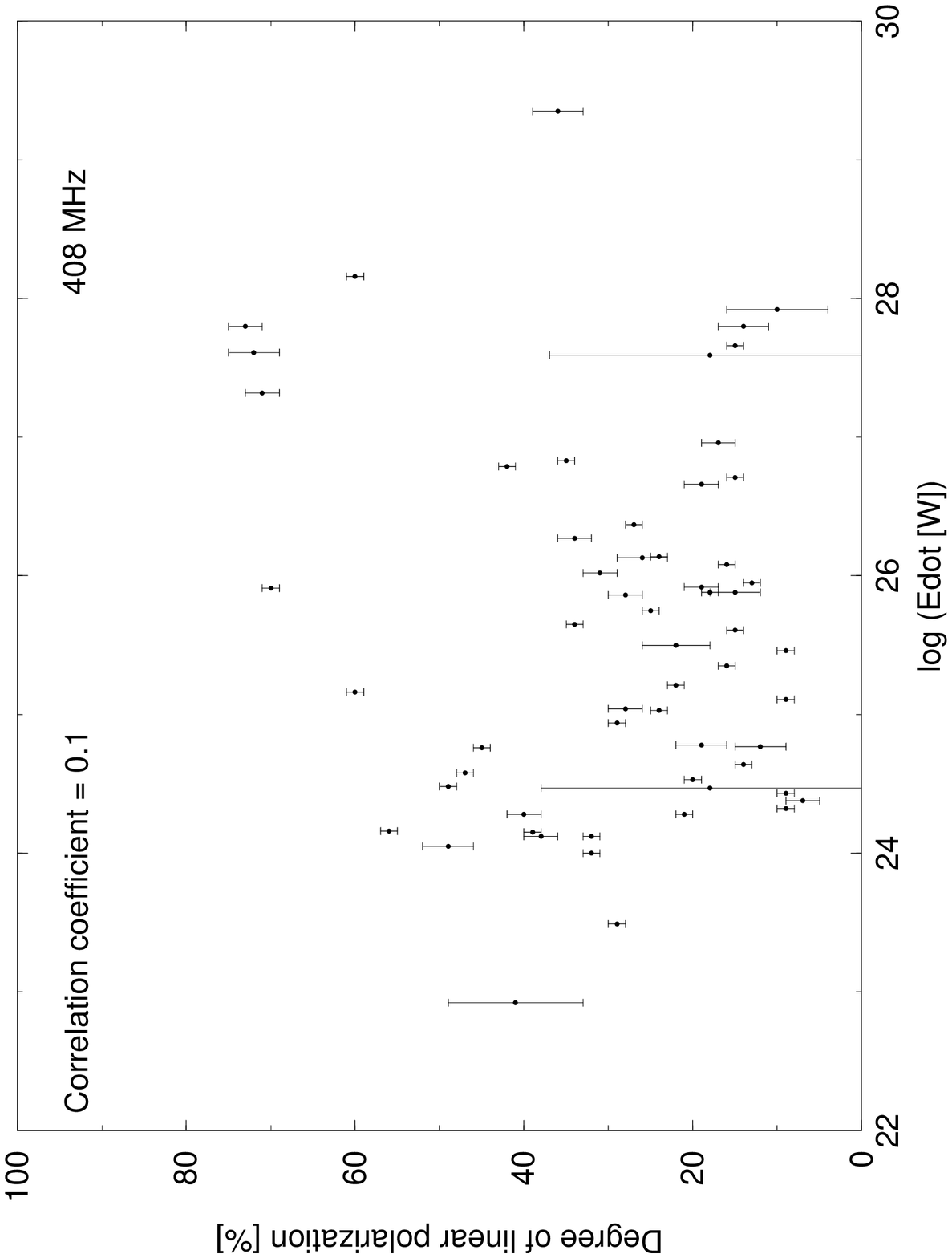}}
\vspace{.7cm}
\epsfysize5cm
\rotate[r]{\epsffile[120 50 520 450]{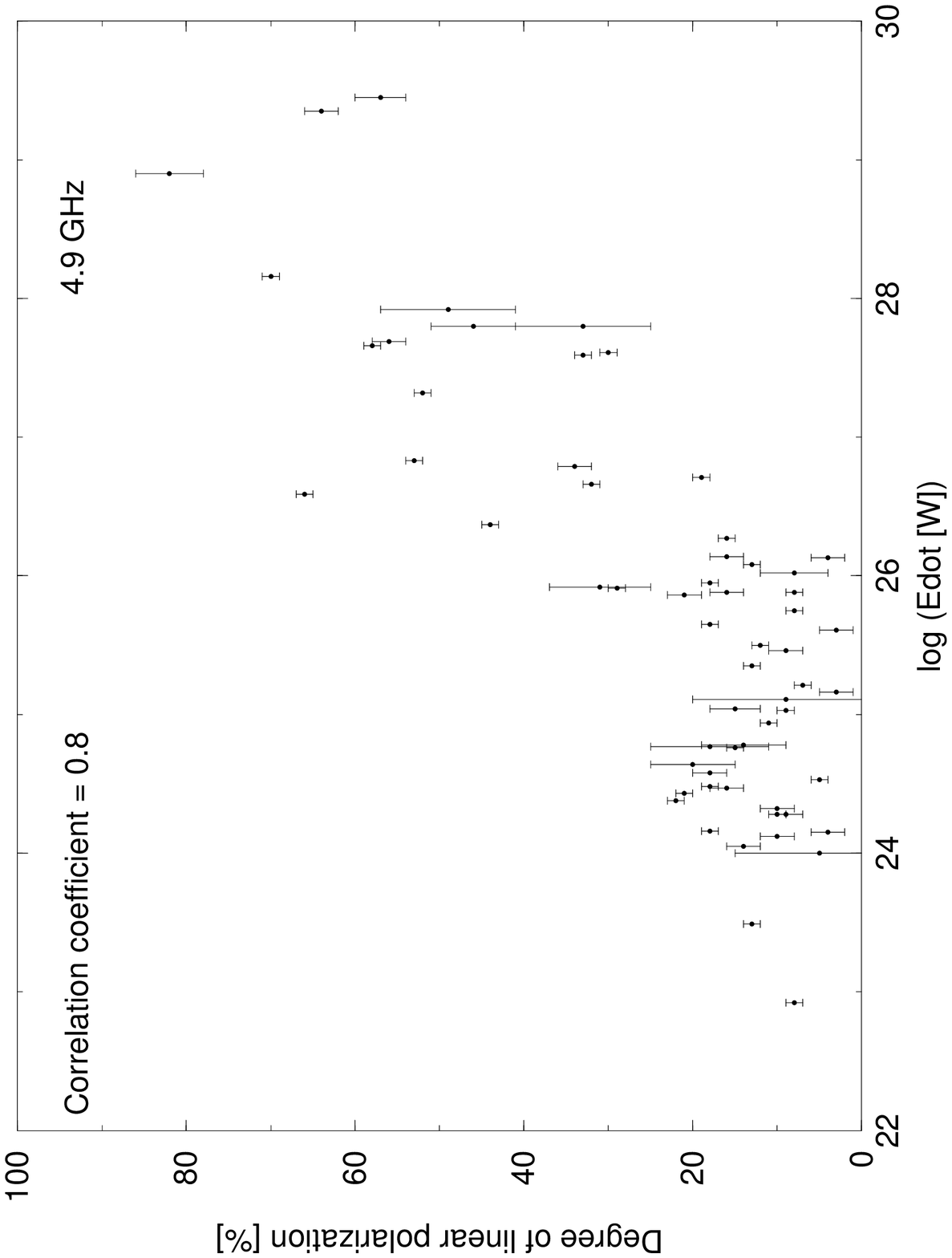}}
\caption[]{Measured degree of linear polarization versus the spin down
luminosity $\dot E$ at 408 MHz (upper plot) from Gould \& Lyne (1998) 
and 4.9 GHz (lower plot), taken with the 100 m Effelsberg radio telescope. 
The two plots consist of (nearly) the same set of pulsars.
These observations correspond to the predictions of 
Fig. \ref{edot}. At the higher frequency, pulsars with a 
low $\dot E$ have a low degree of linear polarization. At low 
frequencies no such correlation
can be observed which again corresponds to Fig. \ref{edot}. At low
frequencies the degree of linear polarization is nearly independent
of $\dot E$.}
\label{edotdata}
\end{figure}

As stated in Sect. \ref{intro}, a correlation between the degree of
polarization and the spin-down luminosity $\dot E$ has been observed
at high frequencies. 
\begin{equation}
\dot E=4\pi^2\cdot I \frac{\dot P}{P^3}\ ,
\end{equation}
$I$ is the moment of inertia.
In order to perform our calculations for different
values of $\dot E$, we keep all parameters fixed and change only the
time derivative of the period $\dot P$. As before, the period was set
to 0.6 s and for $\dot P$ we used the values $10^{-16.3}$, $10^{-14.6}$
and $10^{-12.1}$. This corresponds to $\dot E$ values of $9\cdot
10^{23}$ W, $5\cdot 10^{25}$ W and $1\cdot 10^{28}$ W respectively

Figure \ref{edot} shows that for high $\dot E$ pulsars the degree of 
linear polarization decreases at a higher frequency than for low $\dot E$
pulsars. This is in agreement with the observations, especially if one
follows the argument of Sect. \ref{assumptions} and takes the degree
of polarization as an upper limit. At low frequencies, all pulsars
have a high degree of linear polarization as an upper limit. Therefore
no correlation with $\dot E$ is expected. At
higher frequencies -- in contrary -- the upper limit for low--$\dot E$
pulsars is low, whereas high--$\dot E$ pulsars can still
exhibit highly linearly polarized radio emission. A correlation is
therefore expected at those frequencies. 

\subsubsection{Period}

The next parameter we vary is the period. In the list of
observational constraints in Sect. \ref{intro} it was stated that
only for short-period pulsars highly linearly polarized high frequency radio
emission has been observed. We therefore keep all parameters  
fixed and vary
only the period. The emission height was set to 2 \% of the light
cylinder radius (and therefore changes in absolute numbers as the light
cylinder radius varies with the period). 

\begin{figure}
\vspace{.5cm}
\epsfysize5cm
\rotate[r]{\epsffile[120 50 520 450]{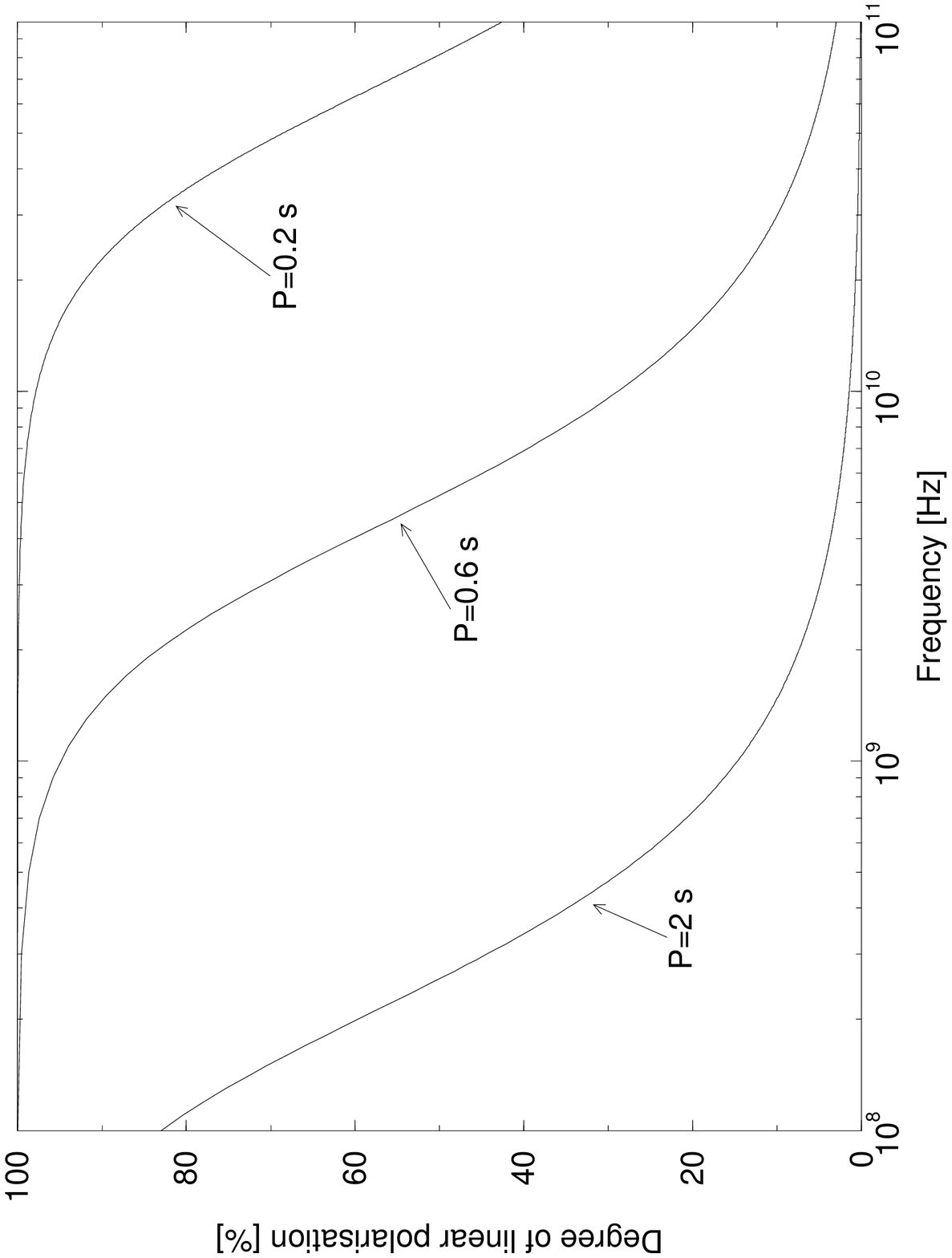}}
\caption[]{Degree of linear polarization of one natural wave mode
versus frequency for different pulsar periods. $\dot P$ was held fixed
at $10^{-14.6}$, the emission height at 2\% and $R_{\rm LP}$ 
at 20 \% of
$R_{\rm LC}$. Short-period pulsars can maintain a higher degree
of polarization up to a higher frequency than their long-period counterparts.
}
\label{period}

\vspace{0.5cm}
\epsfysize5cm
\rotate[r]{\epsffile[120 50 520 450]{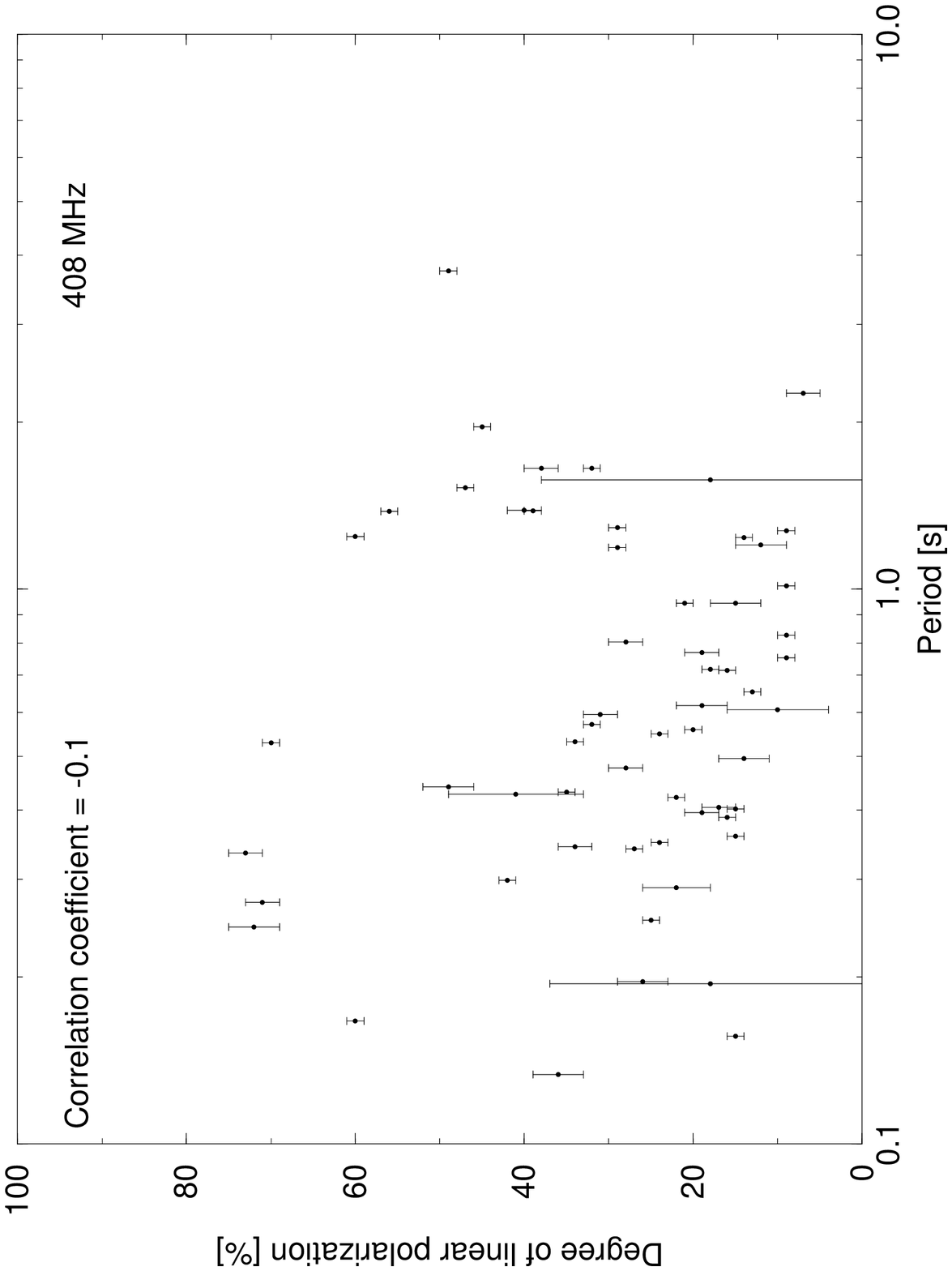}}
\vspace{0.7cm}
\epsfysize5cm
\rotate[r]{\epsffile[120 50 520 450]{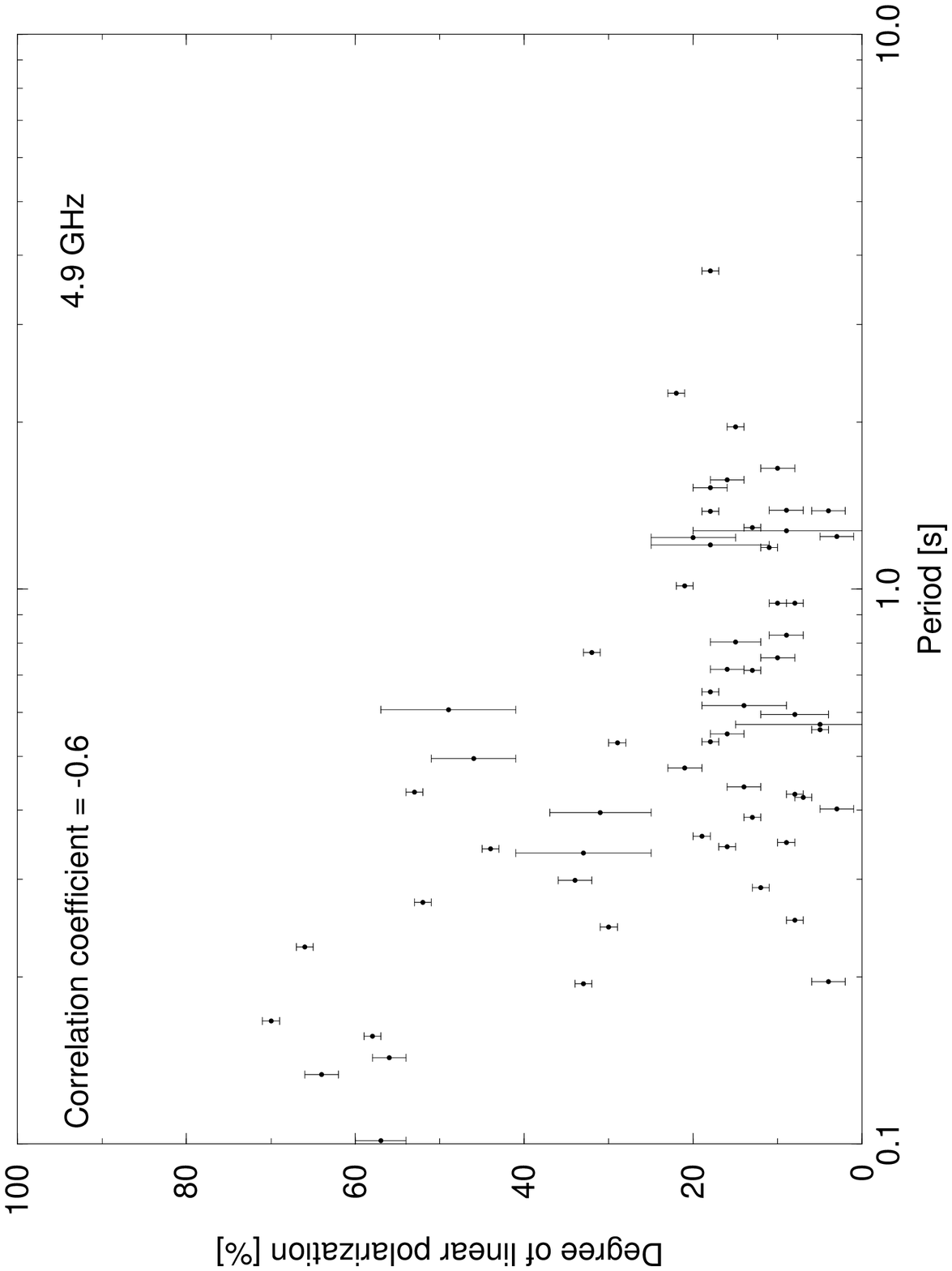}}
\caption[]{Measured degree of linear polarization versus period at 408 MHz 
(upper plot) from Gould \& Lyne (1998) 
and 4.9 GHz (lower plot), taken with the 100 m Effelsberg radio telescope. 
The two plots consist of (nearly) the same set of pulsars.
This observations correspond to the predictions of Fig. \ref{period}. 
At the higher frequency, pulsars with a long period radiate with a low
degree of linear polarization. At low frequencies no such correlation
can be observed which again corresponds to Fig. \ref{period}. At low
frequencies the degree of linear polarization is nearly independent
of the period.}
\label{Pdata}
\end{figure}

Figure \ref{period} shows that short-period pulsars can 
maintain a higher degree
of linear polarization up to a higher frequency than long-period
pulsars. This is in excellent agreement with the observations. At low
frequencies  the linear polarization can be high for
all periods, we therefore would not expect an observable correlation. 
At high frequencies in contrast we therefore {\it do} expect an inverse
correlation. 

\subsubsection{Lorentz-factor of the background plasma}\label{lorentz}

\begin{figure}
\vspace{.5cm}
\epsfysize5cm
\rotate[r]{\epsffile[120 50 520 450]{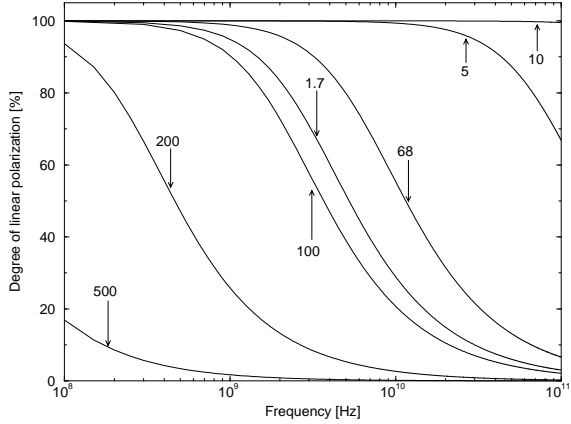}}
\caption[]{Degree of linear polarization of one natural wave mode
versus frequency for different background Lorentz-factors 
(the values are written
next to the curves). As the intersecting angle
between the propagating wave and the magnetic field is Lorentz
transformed, the degree of linear polarization first increases with
$\gamma_{\rm bg}$ until
the transformed angle reaches $90^\circ$ and then decreases again for
larger resulting angles. The difference is the handedness of the
circular contribution for high and low $\gamma_{\rm bg}$ factors.
}
\label{gamma}
\end{figure}

The Lorentz-factor of the background plasma is important because the
intersecting angle gets Lorentz transformed. If one increases
the Lorentz factor at a given point, the Lorentz-transformed
intersection angle first increases up to $90^\circ$ in the plasma frame, 
resulting in fully linearly polarized modes of transverse propagation 
(see Fig. \ref{trans}). 
A further increase of $\gamma_{\rm bg}$ leads to angles
$>90^\circ$. The degree of linear 
polarization then decreases again and the circular contribution 
changes its handedness. This behaviour can be seen in
Fig. \ref{gamma}. One consequence is that if the Lorentz transformed 
intersecting angle between the propagating wave and the magnetic field
is close to perpendicular, small, local variations in the
conditions can result into changes in the handedness of the circular
polarization. As we have to assume that the plasma conditions are
highly inhomogeneous in space and time, this effect could well account
for sudden changes in handedness observed in low degree circular
polarizations. 

An inference which can be made is that the
background Lorentz factor cannot be very large, e.g. larger than a few
hundred. 

\subsection{Refractive indices}

So far we only regarded the polarimetric properties of {\it one} 
natural wave mode. But, as mentioned in Sect. \ref{intro}, 
observations suggest that both modes are present in the pulsar radio 
emission as the familiar OPMs. They can be observed both separated 
(leading to highly polarized parts in the pulse profile with a 
smooth variation
of the PPA) and superposed (leading to weakly polarized parts of the 
profile with often erratic variations in the PPA). 

A spatial separation of polarized modes in a medium is caused 
by birefringence, 
where both modes propagate independently with different indices of 
refraction (a familiar example for such a medium is the calcite crystal).
Different indices of refraction lead to an angular separation which may 
result into an independent observation of the two modes. 
The magnitude of this angular separation is given by 
\begin{equation}
\xi_{\rm refr}=\frac{\delta(n_+-n_-)}{\delta \psi_{\rm S}}
\end{equation}
(\cite{M79}). Clearly the separation 
depends on the difference between the two 
indices of refraction. The larger the difference is, the better is the 
chance, that we observe the two modes separately. If the difference
is small, we expect increasing superposition of the two modes, thus
leading to a depolarization. 

\begin{figure}
\vspace{.5cm}
\epsfysize5cm
\rotate[r]{\epsffile[120 50 520 450]{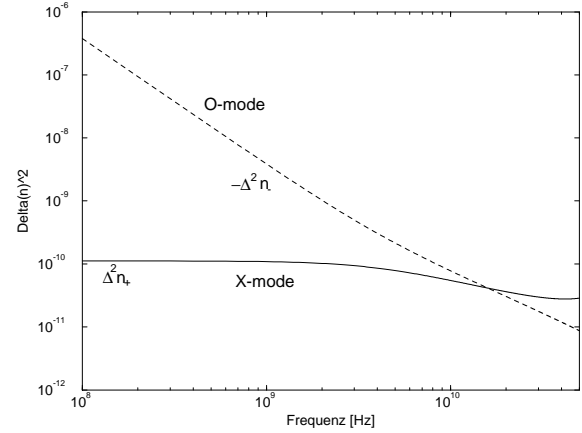}}
\caption[]{Deviations of the two squared indices of refraction of 
the natural wave modes from unity ($n^2=1+\Delta n_\pm^2$). Note that
$\Delta n_-^2$ is negative. $\Delta n_+^2$ (solid line) belongs to the
refractive index of the X--mode and $\Delta n_-^2$ (dashed line) 
to the one of the O--mode. The assumed parameters are those of the 
canonical pulsar defined in Sect. \ref{canonical}. For the O--mode, 
the magnitude of $\Delta n^2$ is much larger and the frequency 
dependence is much 
stronger than for the X--mode. The difference between the two refractive
indices clearly decreases with increasing frequency.
}
\label{brech}
\end{figure}

Figure \ref{brech} shows the frequency dependence of the refractive indices
of the two natural wave modes. The deviations of the indices from unity are 
displayed: $n_\pm^2=1+\Delta n_\pm^2$ (see Eq. \ref{brecheq}). The $+$mode
corresponds to the X-mode (the polarization vector lies in the plane
spanned by $\vec k$ and $\vec B$), the $-$mode to the O-mode 
(polarization vector perpendicular to $\vec k$ and $\vec B$).

Two conclusions can be drawn from the diagram: 
{\bf 1.} The refractive index of the O--mode is much stronger 
influenced by propagation effects than the one of the X-mode. 
This corresponds to the result of Barnard \& Arons (1986), 
that the X--mode can propagate through the magnetosphere nearly 
undisturbed, whereas the O--mode is tight to the bending of the 
field lines. 
{\bf 2.} The difference between the two refractive indices decreases
with increasing frequency. This may lead to increasing
superposition of the two natural wave modes (either intrinsically or
through the limited spatial and temporal resolution of the observation).
If the phases of the modes are uncorrelated, this will cause a 
depolarization of the observed polarization. 

If this conclusion is correct -- depolarization at higher frequencies
through superposition of OPMs -- an additional depolarization envelope
is superposed on our previous discussion. This depolarization could 
be responsible for the low number of pulsars which show the 
increasing degree of circular polarization towards high frequencies.

\begin{figure}
\vspace{.5cm}
\epsfysize5cm
\rotate[r]{\epsffile[120 50 520 450]{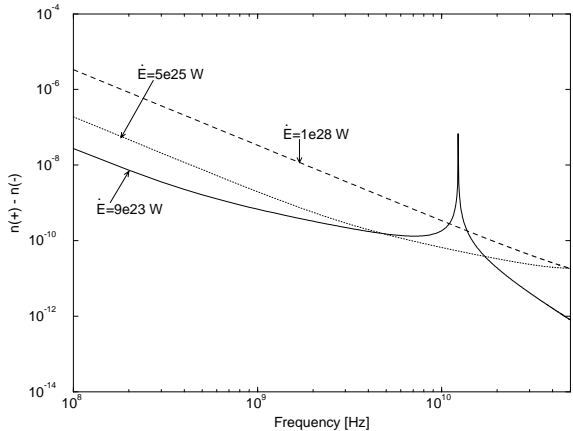}}
\caption[]{The difference between the refractive indices of the 
two natural wave modes $n_+-n_-$ versus frequency for different
values of $\dot E$. The difference is larger for pulsars with a
high $\dot E$ which should therefore suffer depolarization effects at
higher frequencies that low $\dot E$-pulsars. 
The peak in the curve for $\dot E=5\cdot10^{25}$ W is caused by a
resonance when the wave hits the Doppler shifted gyro frequency and
should therefore be neglected (see text).
}
\label{Dbrech}
\end{figure}

The difference between the two indices depends on other pulsar 
parameters as well. Figure \ref{Dbrech} displays the frequency 
dependence of this difference for various values for $\dot E$.
$n_+-n_-$ is larger for larger values of $\dot E$. This leads
to a depolarization at higher frequencies compared to pulsars with a 
low $\dot E$.
The peak in the curve for $\dot E=5\cdot10^{25}$ W is caused by a
resonance when the wave reaches the Doppler shifted gyro frequency. 
It is questionable whether the made approximation is valid at
this particular point. Following Melrose (1979) and Beskin et al. (1993)
we therefore neglect the resonance and join the asymptotic course on 
both sides of the peak.

\section{Conclusions}

We have used the approximation of the dielectric tensor for the low
frequency limit to derive the general properties of the natural wave
modes in the radio pulsar magnetosphere. The approximation of the
dielectric tensor in the given limit implies that the plasma is
sufficiently cold. Following Melrose (1979) this should be true for
pulsar magnetospheres.

Electromagnetic waves, which propagate through a plasma at an oblique
angle to the magnetic field, adopt two orthogonal natural wave
modes with individual indices of refraction. We identify these modes
with the familiar orthogonal polarization modes. It was our aim to
derive the polarimetric properties of these modes and calculate
their dependence on different parameters under special consideration
of the various angles involved.

We regard only one natural mode assuming that we see only one mode at
a time. This assumption is of course not always true, but the
existence of highly polarized emission in single pulses indicates that
at least sometimes it is true. If both modes are recorded simultaneously, it
will lead to a depolarization of the observed radiation. Our
results can therefore be regarded as upper limits for the observed
degree of polarization. In the case of a highly linearly polarized 
natural mode
we cannot predict the degree of linear
polarization one will observe. But if the mode has only a low level
of linear polarization, a low level for
the observed radiation is predicted.

The main uncertainty of our calculations is the radius of limiting 
polarization ($R_{\rm LP}$). 
However, as we are mainly interested in the qualitative behaviour of the
polarization, we have used a fixed value for $R_{\rm LP}$ without
restriction of generality. For a different value of $R_{\rm LP}$ the 
numbers may change but the general behaviour remains. This is even true 
if one moves $R_{\rm LP}$ to its possible extremes: the light cylinder 
radius and just above the emission region. 

Our results are in excellent agreement with the observations: 
{\bf 1.} The
degree of linear polarization decreases towards high frequencies
(Fig. \ref{freq1}). The decrease in linear polarization 
implies an increase in circular polarization (within this
concept). This is in agreement with recent observations which show the
existence of some pulsars which have this peculiar polarimetric property
(\cite{HKK98}).
{\bf 2.} Given a set of specific pulsar parameters, a variation of
$\dot P$ leads to a variation in the spin down luminosity $\dot
E$. Figure \ref{edot} shows that the degree of linear polarization of 
pulsars with a high $\dot E$ decreases at higher frequencies than for
pulsars with a lower $\dot E$. Thus a correlation between the
polarization at high frequencies and $\dot E$ should exist. Such a
correlation has	been indeed observed. 
{\bf 3.} Similar to the above point, we have varied the period,
keeping all other pulsar parameters fixed (Fig. \ref{period}). 
The degree of linear polarization for long period pulsars 
is expected to decrease at lower frequencies than for 
short periods pulsars. This should give an
inverse correlation between the degree of linear polarization at high 
frequencies and the period. Again, such an inverse correlation has been 
observed at a frequency of 5 GHz.

No investigation could be made for the change of the
polarimetric characteristics along the different field lines, which
are observed during a pulse. The reason is that the qualitative change
depends heavily on $R_{\rm LP}$. For future work it
therefore seems to be important to find an estimate for this distance.

The difference between the refractive indices of the two modes
decreases with frequency. As this implies a closer propagation of the
modes, we expect an increasing superposition. This would lead to
a depolarization of the observed radiation at high 
frequencies. Such a depolarization could account for low abundance of
pulsars which show an increasing degree of circular polarization with
frequencies. 

We note again that our calculations are based on the two assumptions
that the background plasma is cold and is dominated by particles 
with one sign of charge. The obvious advantage of these assumptions
-- negligible spread of the energy distribution and no
strong pair production -- is their physical and mathematical
simplicity. The more it is astounding that we can qualitatively
reproduce the complex properties of pulsar radio polarization.
In general both assumptions are not widely accepted. Either the
energy distribution of the plasma is usually taken to be extended
(Arons \& Barnard 1986; Lyutikov 1998 
and references therein) or the secondarily produced electron and positron
pairs are expected to dominate the particle content of the pulsar 
magnetosphere (e.g. Sturrock 1971, Cheng \& Ruderman 1977 and
Daugherty \& Harding 1982).
However the expressiveness of our model deserves further considerations.

\acknowledgements  We want to thank R. Wielebinski, R.T. Gangadhara, 
M. Kramer,  A. Jessner, D.R. Lorimer and K.M. Xilouris
for helpful discussions and support of this work. We also thank the 
anonymous referee for helpful comments on the manuscript.

\end{document}